# Planar Hall effect in $Y_3Fe_5O_{12}$ (YIG)/IrMn films

X. Zhang[1]

[1] Beijing National Laboratory for Condensed Matter Physics, Institute of Physics, Chinese Academy of Sciences, Beijing 100190, China

Correspondence and requests for materials should be addressed to X. Z. (zhangxu1986@iphy.ac.cn)

**ABSTART**

The planar Hall effect of IrMn on an yttrium iron garnet (YIG = $Y_3Fe_5O_{12}$) was measured in the magnetic field rotating in the film plane. The magnetic field angular dependence of planar Hall resistance (PHR) has been observed in YIG/IrMn bilayer at different temperatures, while the GGG/IrMn (GGG= $Gd_3Ga_5O_{12}$) shows constant PHR for different magnetic field angles at both 10 K and 300 K. This provides evidence that IrMn has interfacial spins which can be led by FM in YIG/IrMn structure. A hysteresis can be observed in PHR-magnetic field angle loop of YIG/IrMn films at 10 K, indicating the irreversible switching of IrMn interfacial spins at low temperature.

## I. INTRODUCTION

The coupling between a ferromagnetic (FM)-antiferromagnetic (AFM) bilayer at the interface can lead to a shift of the hysteresis loop along the magnetic field axis due to unidirectional anisotropy. This phenomenon is referred as exchange bias (EB), which often accompanies with the enhancement of coercivity. Due to the pinning effect of magnetic moment in the adjacent FM layer, exchange bias effect has become an integral part in modern spintronics devices such as giant magnetoresistive heads[1].

Although EB was discovered more than 60 years ago[2], the mechanism of it is still a controversial and attracting object to be comprehended as one fundamental physical issue. To understand the mechanism of EB effect, much effort has been dedicated to investigate the behaviors of FM layer in EB systems,[3-5] while the attention on the essential feature of AFM is relatively less due to the notorious difficulties of requirement of large-scale facilities, such as neutron diffraction or XMCD. The theory has predicted that the unidirectional anisotropy in EB system derives from the interfacial uncompensated AFM spin, which does not reverse with the magnetization reversal of the FM layer. However, there is still no solid evidence to prove the existence of it even in the intensively studied Co (-Fe)/IrMn (111) in-plane exchange-biased system.[6-8] Compared to the measuring methods above, there is a more common way to investigate EB system, which is through the electronic transport properties. In the past years, EB effect measured by planar Hall effect has become a more interesting subject due to higher signal-to-noise ratio compared with the magnetoresistance or spin valve configuration.[3,9,10]

In a FM metallic film, the magnetoresistance can be separated into longitudinal and transverse part due to the anisotropic scattering of conduction elections. For a single domain model, the electric field can be described as:[11]

$$E_x = j\rho_\perp + j(\rho_\parallel - \rho_\perp)\cos^2\alpha \tag{1}$$

$$E_y = j(\rho_\parallel - \rho_\perp)\sin\alpha\cos\alpha \tag{2}$$

where the current density j is assumed along the x-axis direction, the magnetization of the single domain is at angle α with respect to x-axis, while $\rho_\parallel$ and $\rho_\perp$ are the resistivity parallel and perpendicular to the magnetization, respectively. Equ. (1) is for anisotropic magnetoresistance (AMR), while Equ. (2) is for planar Hall effect (PHE).

By using AMR effect and PHE, Guohong Li .etc have performed significant work on investigating the NiFe/NiMn EB multilayer and spin-valve.[10] And recently, researches have shown that the reversal of AFM magnetization can lead to the AMR effect in the tunnel junction structures.[12-14] One significant work is the observation of large AFM-tunneling anisotropic magnetoresistance (TAMR) in the MTJs based on NiFe/IrMn/MgO/Pt structure by B.G. Park group,[12] which emphasizes the significant role of AFM in generating the TAMR.

However, these works mentioned above contains both conductive FM and AFM, suggesting that the magnetoresistance signal is consisted of the contribution from both layers. In this condition, the FM may provide an important part of the AMR signal since the reversing of its magnetization can cause a large variation of magnetoresistance. This prevents us from directly observing the magnetization state of AFM layer and the proportion of AMR contributed by AFM is also not clear. In spite of this, these works shed the light on studying the AFM properties in EB system by AMR and PHE. In this work, in order to study the behavior of AFM with other disturbance excluded, an almost isolated FM material $Y_3Fe_5O_{12}$ (YIG) was chosen as FM layer. Through this way, all the magnetoresistance signal would derive from IrMn, thus the result would give out the more specific and directly information of AFM IrMn. By using isolated FM layer in the FM-AFM structure, it provides evidence that there exists uncompensated spins at the interface of IrMn that can be reversed by FM in the YIG/IrMn system at room temperature, and the anisotropy of AFM interfacial spins get enhanced down to the low temperature region.

## II. Experimental

In this study, the YIG substrates is consist of 4 μm thick single crystalline (111) YIG layers grown by liquid phase epitaxy on (111) $Gd_3Ga_5O_{12}$ (GGG) substrates, and single crystalline (111) GGG substrates are also used in this work. Films with structure of YIG/$Ir_{25}Mn_{75}$ (1.8-15 nm)/Ta (2 nm) and GGG/$Ir_{25}Mn_{75}$ (5 nm)/Ta (2 nm) were deposited by DC magnetron sputtering, where (111) texture of the substrates is to promote (111) texture in $Ir_{25}Mn_{75}$ layer, and Ta serves as a cap layer to protect the sample from oxidation/contamination. The vacuum of the sputtering system was better than $4\times10^{-5}$ Pa, and the working Ar pressure was 0.5 Pa. Sets of up to 18 samples were prepared at a time. Composite Ir-Mn target by placing Mn chips symmetrically on the Ir target was used to deposit $Ir_{25}Mn_{75}$ film. The composition was determined by inductively coupled plasma-atomic emission spectroscopy (ICP-AES). The samples were patterned into four terminal Hall bars. For simplicity, $Ir_{25}Mn_{75}$ hereinafter is denoted as IrMn. M-H, PHR–θ and PHR-H curves were measured by a vibrating sample magnetometer (VSM) and physical property measurement system (PPMS) at different temperatures, respectively, where PHR is planar Hall resistivity and θ is the external magnetic field angle.

## III. RESULTS AND DISCUSSION

Magnetic hysteresis loops measured by VSM shows that the YIG films are magnetically soft and isotropic in the film plane. As shown in the upper panel of Fig. 1(a), at room temperature, the YIG film has an in-plane saturation field of about 60 Oe, and the loop remains the same as the sample rotates 90° in the same plane (not shown). The YIG film has a spontaneous magnetization ($4\pi M_S$) of 1.73 kG, which can be obtained in the out-of-plane loop due to the shape anisotropy, as indicated in the bottom panel of Fig. 1(a). These magnetic properties is in agreement with the YIG films reported before.[15] Fig. 1(b) shows the hysteresis loop of a typical exchange bias structure of CoFe (8 nm)/IrMn (10 nm), which displays a saturation field smaller than 300 Oe and an exchange bias field of about 120 Oe, proving that the IrMn deposited by our composite target is reliable antiferromagnetic material. Fig. 1(c) and (d) exhibit the magnetic field angular dependence of PHR for mono-CoFe (8 nm) layer and CoFe (8 nm)/IrMn (10 nm), respectively. In the angular-dependent measurement, the sample was rotated first from θ=0° to 360° and backwards from to θ=360° to 0° in a fixed magnetic field. The data in Fig. 1 (c) was measured under 2 kOe magnetic field at 300 K, which is larger than the saturation field of CoFe. The behavior of CoFe layer illustrates that the PHR of single FM layer would exhibit a sinusoidal dependence of magnetic field angular with period of 180°, and also no difference between the clockwise and anticlockwise curve is observed. Note that the curve is not symmetric of zero resistance point as expected for uniaxial anisotropy of FM materials, which can be explained by the misalignment of the Hall bar that measures the voltage, leading to an additional resistance along the current direction. When coupled to IrMn, the CoFe (8 nm)/IrMn (10 nm) (referred as CoFe/IrMn hereinafter) structure develops exchange bias effect due to the coupling between FM and AFM layer, and an interfacial unidirectional anisotropy would be induced into the system as a result. Fig. 1(b) shows the hysteresis loop of CoFe/IrMn, which displays a saturation field smaller than 300 Oe and an exchange bias field of about 120 Oe. In Fig. 1(d), for the magnetic field below 300 Oe, the angular-dependent curve of CoFe/IrMn shows a distortion compared to mono-CoFe layer data. Since 300 Oe is larger than the saturation field of CoFe/IrMn structure based on the loops in Fig. 1(b), the PHR-theta curve of 300 Oe indicates that the unidirectional anisotropy would lead to a magnetic moment rotating processes different to that of uniaxial anisotropy. When magnetic field increases to 2 kOe, which is much larger than the saturation field obtained in Fig. 1(b), although no obvious distortion can be observed in the PHR curve, there is still a slightly deviation when fitting PHR-$\theta$ data by a sine function. This suggests that most of the CoFe layer synchronously rotates with the external magnetic field of 2 kOe. It is worth mentioning that because of the coupling between CoFe and IrMn, some of the interfacial IrMn spins should rotates certain degree with CoFe, which also should have contribution to PHR. However, owning to the much larger signal of 8 nm thick CoFe layer, this PHR from IrMn can hardly be confirmed. In addition, it is worth noting that the PHR-$\theta$ curve under 50 Oe field shows a hysteresis behavior. Since 50 Oe is not large enough to reverse the CoFe spins pinned by IrMn, and also PHR-$\theta$ curve at this field shows a smaller amplitude when compared with the curves of 300 Oe and 2 kOe, which suggests that the FM moments still rotates a small angle and this spin flopping follows a different route in the clockwise and anticlockwise magnetic field cycling procedure, i.e. the FM moment shows an irreversible switching under 50 Oe field. Recently, several researches have reported the AMR derives from AFM in the FM-AFM EB systems, proposing a new possibility to obtain large magnetoresistance in relatively small fields.[12-14] However, these studies probed AFM with conductive FM layer, which would lead to difficulty of distinguishing AFM signal from FM signal, thus the specific behavior of AFM is still not clearly revealed.

In order to investigate the interfacial magnetization of AFM in an FM-AFM coupled system more independently, antiferromagnetic material IrMn was deposit on the isolated ferromagnetic material YIG. Figure 2 shows the angular dependence of PHE for YIG/IrMn (1.8 nm) and (5 nm) at different temperatures. The sample was firstly field cooling to 10 K under a constant field of 3 kOe in the same direction as the magnetic field applied during the film growth, and then measured at increasing temperatures in the magnetic field of 2 kOe, which is larger than the saturation field of common exchange bias system. At each temperature, the measurements were also performed form 0° to 360° and cycling back to 0° for both samples. Clearly, magnetic field angle dependence of PHR can be observed in both YIG/IrMn samples from 10 K to 300 K, indicating an anisotropic magnetoresistance. Since 2 kOe is not large enough to rotate bulk magnetic moments of IrMn as well as YIG is a nearly isolated material, the anisotropy signal can only derive from the rotation of IrMn interfacial moment, confirming that there are uncompensated spins existing in the interface of IrMn that can be rotated by a small magnetic field.

In Fig. 2(a) and (e), a hysteresis of magnetic field angle can be observed for both samples at 10 K, suggesting that the IrMn interfacial moment cannot be fully reversed by the magnetic field of 2 kOe at this temperature. It was reported that in an EB structure of $Co/YMnO_3$ with insulated $YMnO_3$ as antiferromagnetic layer, a similar hysteresis can be observed in the angular-dependent magnetic field measurement of Co at low temperatures[5], where EB effect is more remarkable. However, in contrast to $Co/YMnO_3$ system, the contributions to PHE are all from the AFM layer in YIG/IrMn film, which indicates that the coupled spins of AFM may exhibit the same behavior as adjacent FM spins. Furthermore, according to CoFe/IrMn in Fig. 1(c) and $Co/YMnO_3$ results, this irreversible of IrMn interfacial magnetization implies a combination of uniaxial and unidirectional anisotropy in IrMn interfacial moment at low temperatures. On the other hand, at higher temperatures, the hysteresis and distortion gradually disappear, and the curves are sinusoidal at 300 K, indicating that the IrMn interfacial moment is uniaxial at room temperature. Thus, it can be deduced that the unidirectional anisotropy is the cause of not-fully reversed moment at 10 K, which is analogous to the CoFe/IrMn-50 Oe curve in Fig. 1(d). Similar to the behavior of NiFe/IrMn stack,[13] the fast loss of the unidirectional anisotropy is probably due to the decrease of IrMn anisotropy as temperature approaches to the Neel temperature of IrMn.[16] Therefore, it can be conjectured that the uncompensated spins at IrMn interface has relatively large coupling with IrMn spins around it at low temperature, which leads to the unidirectional anisotropy. When the temperature increases, this exchange interaction becomes weak and it vanishes at room temperature.

Due to the coupling interaction between FM and AFM, the rotation of interfacial magnetic moment is supposed to be led or promoted by FM moments. To verify this scenario, the angular dependence of GGG/IrMn (5 nm) sample was papered, where GGG is a non-magnetic insulated substrate of (111) $Gd_3Ga_5O_{12}$. As shown in Figure 2(i) and (j), the sample with GGG substrate shows isotropy of PHR for the field below 2 kOe at both 10 K and 300 K, this proves that the promotion of FM to the reversing of interfacial moment of AFM. Therefore, the sinusoidal curves of YIG/IrMn samples at 300 K suggests that the coupling between of YIG and IrMn still exists at room temperature, leading to the rotation of some IrMn interfacial spins with uniaxial anisotropy behavior. Besides, in Fig. 2(d) and (h), the curves of 200 Oe are almost identical to those measured under 2 kOe field, indicating that both IrMn-1.8 nm and IrMn-5 nm samples have reached saturation under field as low as 200 Oe. This phenomenon also suggests a very small anisotropy of interfacial IrMn moments, which is different to the large unidirectional anisotropy observed in CoFe/IrMn at low magnetic field (Fig. 1(c)). Whereas there is no solid evidence to explain this contradiction, one possible assumption is that the roughness of IrMn interfaces may be the reason for this difference. Owning to the fact that the series of YIG/IrMn samples are not in-situ grown, which could induce defects and impurities at the interface to cause the interface of YIG become much rougher than that of CoFe. Therefore, a small part of uncompensated spins of the net magnetic moment at IrMn interface may has relatively weak coupling with the bulk than the others due to the roughness of the interface, resulting in the presentation of a smaller anisotropy, these spins are referred as free spins (F spins). At low temperature, the F spins exhibit unidirectional anisotropy due to their exchange interaction of the bulk IrMn, which decreases at room temperature, leading to the YIG/IrMn sinusoidal dependence at 300 K. And the rest of uncompensated pinned spins (P spins) of IrMn net magnetic moment still has strong coupling with IrMn since the Neel (blocking) temperature is much higher than the room temperature, these P spins cannot be rotated or can only be rotated for a small angle under 2 kOe field. And because CoFe/IrMn is grown in-situ, the roughness of the CoFe/IrMn interface should be much smaller than YIG/IrMn interface. Thus the EB effect of CoFe/IrMn are mostly developed by the coupling between P spins and FM layer, which provides the unidirectional anisotropy for CoFe/IrMn at room temperature.

Another evident phenomenon is that both the IrMn-1.8 nm and IrMn-5 nm samples show dramatic deduction at 90 K and 130 K, as well as the overall trend of low field curve phase shifts 90° as the temperature increases, indicating the shift of easy axis. This phenomenon can be explained by the existence of strongly coupled $Fe^{3+}$ and weakly coupled $Y^{3+}$ in YIG, of which $Y^{3+}$ and the ferrimagnetic component of $Fe^{3+}$ magnetic moments are antiparallel to each other[17]. Because the interactions between $Y^{3+}$ ions are weak, $Y^{3+}$ shows paramagnetic properties in the exchange field generated by $Fe^{3+}$ spins. In low temperature region, the total magnetic moments parallel to the $Y^{3+}$ ions due to its large magnetic moments. As the temperature increases, the magnetic moments of the sub-lattice of $Y^{3+}$ quickly decrease, leading to the domination of $Fe^{3+}$ magnetic moments at high temperature. Therefore, the decreasing of PHR at 90 K and 130 K is probably as a result of the competition of the two kinds of moments, and the change of easy axis is because the exchange interaction only exits between $Fe^{3+}$ in the high temperature region.

Additionally, as shown in Figure 2, apart from the larger amplitude of PHR, the YIG/IrMn

(1.8 nm) sample shows basically the same behavior of the YIG/IrMn (5 nm) at different temperatures. Since the anisotropy of AFM decreases with decreasing thickness in ultrathin AFM film,[18] it can be argued that the spins in IrMn is easier to be reversed by FM, which leads to a larger PHE of YIG/IrMn (1.8 nm). By assuming the interface of IrMn is a single domain film structure, this observed angular dependence of the PHR at 10 K can be described by

$$PHR = R_0 + R_1 \cos^2(\varphi_M - \varphi_j) + R_2 \sin[2(\varphi_M - \varphi_j)] + R_3 \cos(\varphi_M - \varphi_j) \qquad (3)$$

where $\varphi_M$ and $\varphi_j$ are the angles of IrMn interfacial magnetization and current direction relative to the sample position of zero degree, respectively. The first and second term represent the ARM caused by the misaligning of Hall bar which leads to a voltage difference along the current direction, while the third and fourth term are classic PHR and an extra PHR generating from unidirectional anisotropy, respectively. The curve of IrMn-1.8 nm and IrMn-5 nm fitted by Eq. (3) is shown in Figure 3, exhibiting good agreement with the experiment data. The parameters obtained from fitting are shown in Table. I. It should be noted that $R_3$ become smaller as the temperature increases, which consists with the decreasing tendency of unidirectional anisotropy observed in Figure 2.

TABLE I. Fitting parameters for YIG/IrMn (1.8 nm) and YIG/IrMn (5 nm) film by Equ. (3) at 10 K and 300 K

| Sample | Temperature (K) | $R_0$ (ohm) | $R_1$ (ohm) | $R_2$ (ohm) | $R_3$ (ohm) |
|---|---|---|---|---|---|
| IrMn-1.8nm | 10 | 5.91278 | $3.93735 \times 10^{-11}$ | 0.00413 | 0.00101 |
| | 300 | 4.94969 | $3.023 \times 10^{-10}$ | 0.00541 | 0.00013 |
| IrMn-5nm | 10 | 4.3297 | $-5.47 \times 10^{-4}$ | -0.00171 | 0.000469 |
| | 300 | 3.97303 | $1 \times 10^{-5}$ | 0.00361 | $6.09181 \times 10^{-5}$ |

A further investigation on the IrMn thickness dependence of Hall angle is summarized in Fig. 4(a). The measurement was performed at 300 K under 200 Oe, and $\Delta\rho_{xy}$ and the $\rho_{xx}$ was obtained by the PHR-H curve while the angle between magnetic field and current is 45°. The PHR-H loops of representative samples, IrMn-1.8 nm and IrMn-5 nm are shown in Fig. 4(b) and (c). Despite the IrMn-1.8 nm sample shows a larger PHR amplitude than IrMn-5 nm sample in Figure 2, it can be observed in Fig. 4(a) that the Hall angle almost remains constant when IrMn is not thicker than 5 nm. When the thickness of IrMn is larger than 5 nm, the Hall angle begin to decrease dramatically and reaches a constant as IrMn is thicker than 10 nm. This confirms the interface nature of PHE in IrMn, and this interfacial effect does not quickly decrease until IrMn is not thick enough.

## IX. CONCLUSION

In conclusion, the PHR of YIG/IrMn samples with different IrMn thicknesses were investigated in this work. The magnetic field angle dependence of PHR can be observed from 10 K to 300 K below 2 kOe, suggesting the existence of uncompensated spins at YIG/IrMn interface. And also a hysteresis derived from unidirectional anisotropy can be observed in the PHR-θ curves at 10 K in YIG/IrMn films. By comparing with GGG/IrMn sample, which PHR-θ curve shows isotropic behavior at both 10 K and 300 K, it can be considered that the interfacial IrMn spins are led by the magnetic moment of YIG.

## ACKNOWLEDGEMENTS

This work was supported by the National Key Basic Research Program of China under Grant No. 2014CB921002, and the National Natural Science Foundation of China under Grant Nos.51171205 and 11374349.

**Figure captions:**

FiG. 1. (a) The in-plane (top panel) and out-of-plane (bottom panel) hysterias loops of YIG, and (b) the in-plane hysterias loop of CoFe (8 nm)/IrMn (10 nm), and (c) the PHR-θ curve of CoFe measured under 2 kOe magnetic field at room temperature, and (d) the PHR-θ of CoFe (8 nm)/IrMn (10 nm) measured under different magnetic fields at room temperature.

FIG. 2. PHR of (a-d) YIG/IrMn (1.8 nm) and (e-h) YIG/IrMn (5 nm) sample when rotating in the applied magnetic field of 2 kOe at different temperatures. The measurements were also performed under 200 Oe for both YIG/IrMn samples at 300 K. (i) and (j) are the magnetic field angular dependence of GGG/IrMn (5 nm) film at 10 K and 300 K, respectively. At each temperature, the PHR-θ curves were measured under different magnetic fields of 200 Oe and 2 kOe, both of which exhibit no obvious variation with magnetic field angular θ.

FIG. 3. PHR-θ cure and its fitting curve of YIG/IrMn (1.8 nm) sample at (a) 10 K and (b) 300 K, and PHR-θ cure and its fitting curve for YIG/IrMn (5 nm) sample at (c) 10 K and (d) 300 K. The fitting curve uses the data of magnetic field clockwisely rotating from 0° to 360°. The data of 10 K and 300 K are fitted by Equ. (3).

FIG. 4.  (a) The IrMn thickness dependence of Hall angle, and the PHR-H curve for (b) YIG/IrMn (1.8 nm) and (c) YIG/IrMn (5 nm) measured at 300 K. The angle between magnetic field and current is 45°.

**Figures:**

FIG.1.

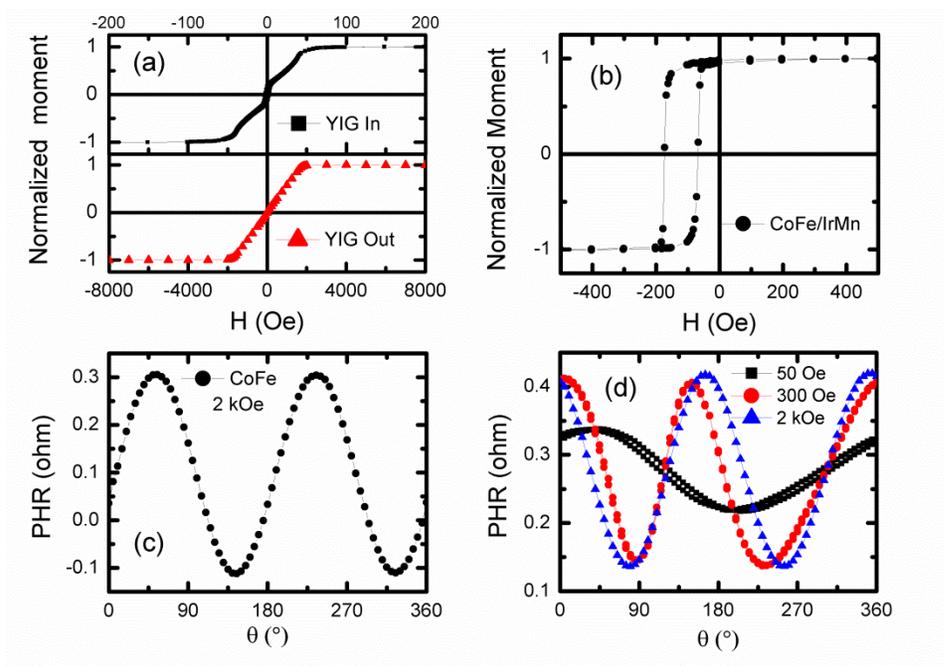

FIG. 2.

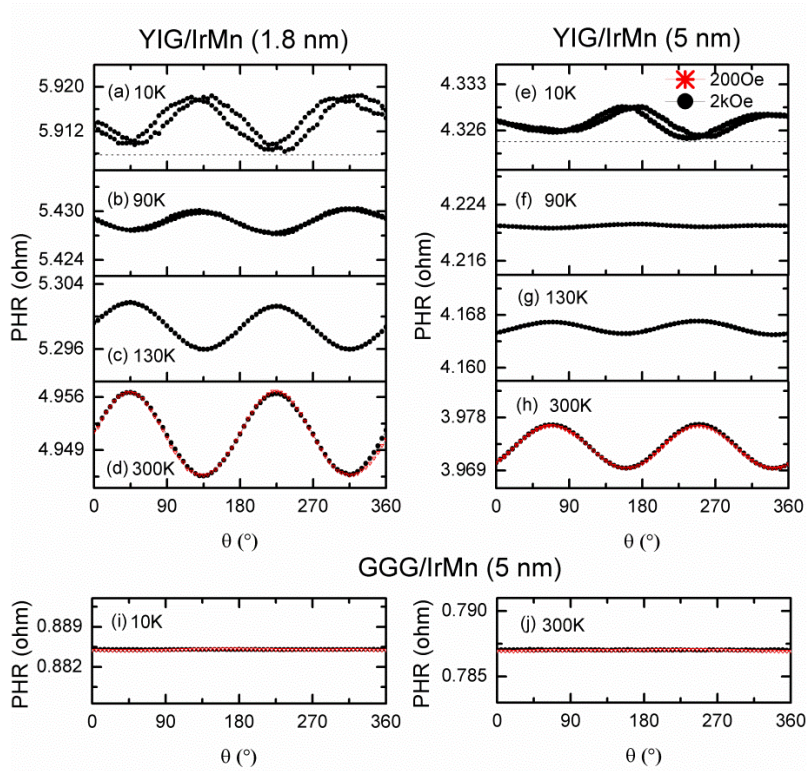

FIG. 3.

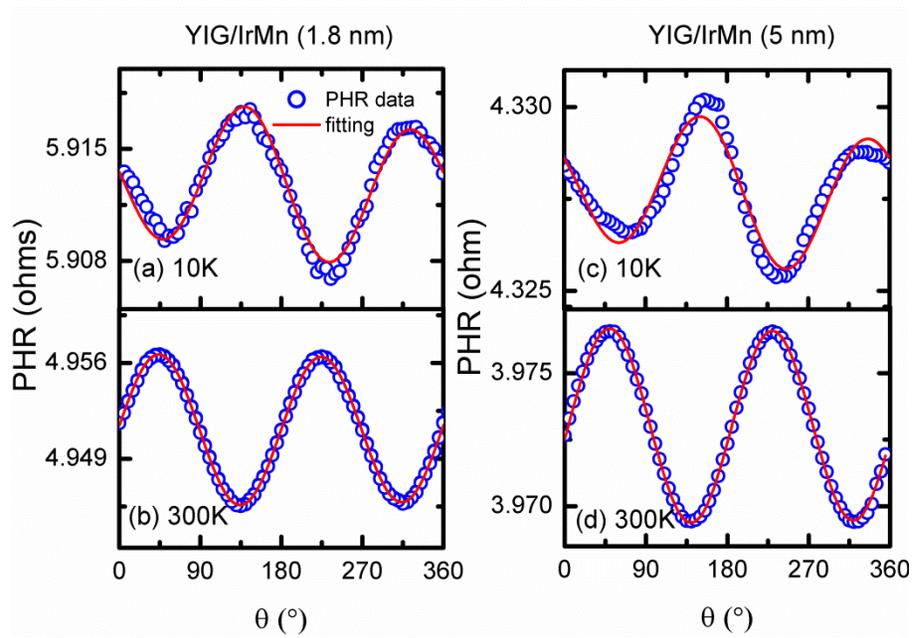

FIG. 4.

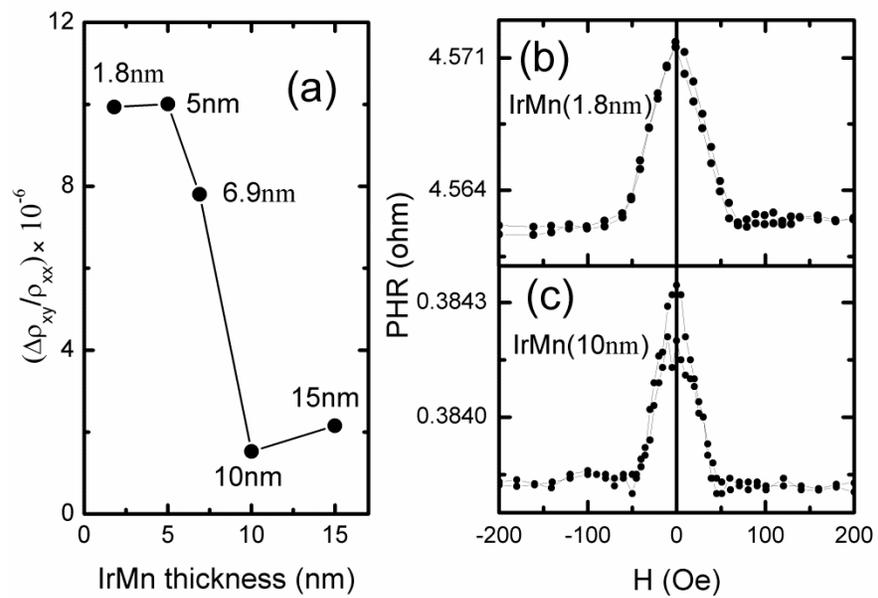